# A Monte Carlo Simulation Study of L-band Emission upon Gamma Radiolysis of Water

K A Pradeep Kumar[1], G A Shanmugha Sundaram[1], S Venkatesh[2], R Thiruvengadathan[1,*]

[1]SIERS Research Laboratory, Department of Electronics and Communication Engineering, Amrita School of Engineering, Coimbatore, Amrita Vishwa Vidyapeetham, INDIA.
[2]Formerly with Department of Sciences, Amrita School of Engineering, Coimbatore, Amrita Vishwa Vidyapeetham, INDIA.
*Corresponding Author's Email: t_rajagopalan@cb.amrita.edu

**Abstract**: Several studies have confirmed visible light and ultraviolet emission during water molecule radiolysis; however, radiofrequency (RF) emissions have been scarcely investigated. This simulation study has revealed that the gamma radiolysis of water creates excited hydrogen atoms which emit radio recombination L-band (1 GHz – 2 GHz) radio waves of sufficient strength that a radio-imaging device can detect. The physical and physicochemical stages of radiolysis of water have been modeled via application of Monte Carlo simulation techniques up to 1 ps from the onset of gamma photon interaction with water molecules. Subsequently an L-band emission analysis up to a few ms duration was accomplished to obtain and understand RF power emission characteristics as a function of gamma radiation dose rate. The generated L-band RF power is $10^{-21}$ W per 1.17 MeV gamma photon. The generated RF power varies linearly with absorbed dose rate up to 1.13 Gy/h, beyond which non-linearity sets in due to Stark broadening. The emission has a positive radio spectral index of 1.738, confirming that the emission mechanism is spontaneous radio recombination.

# 1 INTRODUCTION

It has been well established that an excited hydrogen atom undergoes spontaneous decay emitting electromagnetic radiation in the ultraviolet (UV) – Lyman series, visible – part of Balmer series, and radiofrequency (RF) spectrum [1]–[3]. Further, gamma radiolysis of water creates highly excited hydrogen atoms, among other radiolytic species [4] [5]. However, does the excited hydrogen atoms produced during water radiolysis emit electromagnetic waves? Indeed UV and visible emissions from hydrogen atoms created during the radiolysis of water have been well documented [6], [7]. At the same time, only a very few research works report the study of RF emission and confine their studies to ground-state spin transition known as hyperfine structure transition (HFST) emission at 1.4 GHz. Rajendran et al. briefly discuss HFST RF emission due to water radiolysis [8]. Kolotkov et al. and Yakubov et al. provide a qualitative treatment of RF emission due to water radiolysis lacking a detailed analysis of the emission mechanism [9], [10].

The lack of in-depth investigations of RF emissions during water radiolysis could be attributed to the reasons stated below. While UV and visible emissions of a hydrogen atom are caused due to allowed transitions that have very short decay times, 1.4 GHz HFST RF emission has very low transition probability and ultra-slow decay time [1]. For example, the decay time for the 1.4 GHz HFST is 10 million years [1]. Hence, about 10 million hydrogen atoms are required to achieve an emission rate of approximately one RF photon per year. Besides, amongst the different radiolytic species, the hydrogen atom yield is one of the lowest [11]. Hence it is understandable that the HFST could not produce RF emissions with sufficient signal strength during radiolysis of water.

Besides HFST emission, excited hydrogen atoms can also emit radio waves known as radio recombination lines (RRL) upon relaxation to certain lower energy states. However, unlike the hydrogen HFST, RRL transitions are allowed, with a short mean decay time of about a few milliseconds [12]. Nevertheless, a hydrogen atom is highly reactive and tends to recombine quickly [13]. The highly excited hydrogen atom responsible for L-band RRL emission, is called a Rydberg hydrogen atom. The Rydberg hydrogen atoms are relatively large and may be easily ionized before they have a chance of emission [5]. Also, as we know today, both RRL and HFST emission originate from very low-density atomic hydrogen clouds in outer space at extremely high or low temperatures [1]. Naturally, a question is prompted. Do the hydrogen atoms that are produced during gamma radiolysis of water support RRL emission? In this article, we have developed a radio-frequency source model, which could relate the intensity of radio waves emitted during the gamma radiolysis of water to the absorbed gamma dose rate.

The theory of gamma radiolysis of water has been established to the point of tracking various radiolytic species, including atomic hydrogen up to a time scale of a millisecond [14]. Further, the theory of L-band RRL emission from atomic hydrogen has been demonstrated [1], [15]. However, the atomic hydrogen source responsible for L-band RRL emission is not associated with the gamma radiolysis of water [1], [15] . To the best of the authors' knowledge, there has been no study to date that reports L-band RRL emission from atomic hydrogen produced during gamma radiolysis of water. The L-band RRL emission is rather related to the hydrogen atoms in the native form available ubiquitously throughout the universe [1]. Specifically, we do not know the RRL intensity for a specific gamma dosage absorbed by water. To determine the relationship between L-band RRL intensity and the absorbed gamma dose rate, we conducted a systematic study by performing the following tasks: a) Develop a model to distinguish L-band RRL emission among other emissions caused by gamma radiolysis of water. B) Develop a model to simulate RF emission and compute L-band RF power emitted. C) Obtain radio spectral index parameter and validate the modelled RRL emission mechanism. D) Study the impact of broadening aspects on emitted RRL signatures. E) Determine the relationship between the intensity of L-band RRL radio waves emitted during the gamma radiolysis of water and the absorbed gamma dose rate.

## 2   THEORETICAL BASIS FOR L-BAND RF EMISSION UPON GAMMA WATER RADIOLYSIS

A gamma photon travels through a random distance, called free-flight, on its passage through a water body. After the photon travels the free-flight length, it undergoes an interaction with a water molecule. The interaction triggers a series of events, creating several species. The sequence of events, which spans a total time greater than $10^{-3}$ s, are grouped under physical stage ($10^{-18}$ to $10^{-12}$ s), physicochemical stage ($10^{-15}$ s to $10^{-12}$ s), non-homogenous chemical stage ($10^{-12}$ s to $10^{-6}$ s), and homogenous chemical stage ($10^{-6}$ s to $10^{-3}$ s) [4], [16], [17].

The physical stage begins with the interaction of the gamma photon with a water molecule and is associated with the energy transfer to the water molecule. The interaction process can be one of the three physical processes: photoelectric effect, Compton scattering, or pair-production. However, for the interaction of gamma radiation, in the energy range of interest, 100 keV to 1.17 MeV with water, the interaction process is primarily Compton scattering [4]. Hence, in this article, the photoelectric effect and pair production processes are not considered. In Compton scattering, an electron is knocked off from the molecular orbital (MO) of water (Eq. 1). A significant fraction of the incident photon energy is transferred to the electron. The high-energy electron is called Compton electron. Further, the gamma photon invariably undergoes a change in direction (scattered) in both the azimuth and elevation plane after the scattering. The Compton electron does not trace the path of the incident gamma photon but assumes a new direction of travel.

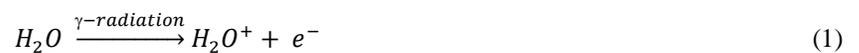

$$H_2O \xrightarrow{\gamma-radiation} H_2O^+ + e^- \qquad (1)$$

A gamma photon's mean free path in liquid water is in the order of few centimeters, while that of the electrons is only a few nanometers (seven orders of magnitude lesser) [17]. Hence the Compton electron ionizes and excites many water molecules in highly localized regions (called spurs and blobs), producing many new electrons termed secondary electrons (Eq. 2,3). The secondary electrons can trigger further ionization and excitation, resulting in an electron shower (Eq. 2).

A water molecule has five filled molecular orbitals (MO). Though all the ionization species have the same chemical formula – $H_2O^+$, the MO from which the electron is knocked off is different, resulting in five different ionized species corresponding to each MO [4], [17]. These ionized species are represented as $1b_1$, $3a_1$, $1b_2$, $2a_1$ and $1a_1$. In an excited species, the ground configuration of the water molecule, which is $(1a_1)^2 (2a_1)^2 (1b_2)^2 (3a_1)^2 (1b_1)^2$, is elevated to various higher energy levels, resulting in different excitation species [18]. The prominent excitation species are $\tilde{A}^1B_1$, $\tilde{B}^1A_1$, and plasmon [18].

$$H_2O + e^- \rightarrow H_2O^+ + 2e^- \quad (2)$$
$$H_2O + e^- \rightarrow H_2O^* + e^- \quad (3)$$

The excitation species ($H_2O^*$), indicated in Eq. (3), further disintegrate as per the reactions specified by Eq. 4 – Eq. 8 [4], marking the onset of the physicochemical stage [16]. At the start of the physicochemical stage, the radiolytic species ($10^{-15}$ s) created are electrons, the ionized species $H_2O^+$, and the excited $H_2O^*$ species [16], [19]. However, the energy level of a majority of the electrons soon falls below the subexcitation level (7.4 eV) and is hydrated, given by Eq. 9 [13]. The final species at the end of the physicochemical stage ($10^{-12}$ s) are the ionized species, hydrated electrons ($e^-_{aq}$), and radical products from the disintegration of $H_2O$, such as $H^\cdot$ and $OH^\cdot$.

$$H_2O^* \rightarrow H_2O \quad (4)$$
$$H_2O^* \rightarrow H^\cdot + OH^\cdot \quad (5)$$
$$H_2O^* \rightarrow H_2 + O \quad (6)$$
$$H_2O^* \rightarrow 2H^\cdot + O \quad (7)$$
$$H_2O^* \rightarrow H_2O^+ + e^- \quad (8)$$
$$e^- + H_2O \rightarrow e^-_{aq} \quad (9)$$

The time-lapse between $10^{-12}$ s and $10^{-6}$ s is called the non-homogeneous chemical stage. During this period, chemical reactions occur within the spur between the colliding species, while the other species escape the spur. Further, the spur expands, overlapping with adjacent spurs triggering even more chemical reactions [13]. At the end of the non-homogenous chemical stage, the spurs overlap and enlarge, ultimately becoming non-distinguishable from its surrounding, terminating the spur reactions.

The G value is the physical quantity popularly used to quantify the number of species produced. It is defined as the number of species created per 100 eV absorption of ionizing radiation [17]. The yield of the radiolytic species at the end of the physicochemical stage is called the initial yield, while that at the end of the non-homogeneous chemical stage is called primary yield [13]. Of all the species, we are interested in specifically the temporal yields of the hydrogen atoms ($H^\cdot$) and hydrated electrons ($e^-_{aq}$). Hydrogen atoms are potential RF emitters, while the hydrated electrons influence the spectral broadening of the RF emission lines. The temporal evolution of $H^\cdot$ and $e^-_{aq}$ under gamma radiolysis of deaerated water at 25 °C is shown in **Figure 1** [14]. While the yield of H· is almost constant during the non-homogenous chemical stage ($10^{-12}$ s to $10^{-6}$ s), hydrated electron yield steadily drops from about 4.3 to 2.4.

The homogenous chemical stage occurs from $10^{-6}$ s to $10^{-3}$ s, when the spur reactions have stopped. During this stage, the yield of the hydrated electron drops at a steeper rate, while the rate of yield of hydrogen atoms increases [14].

The electron in a hydrogen atom (H·) can be excited to a very large principal quantum number level (n), such as n greater than 40, during water radiolysis [5]. When these atoms decay from n+1 to n principal quantum level (indicated by Hnα), with n ranging from 149 to 186, it could be verified using the Planck-Einstein relation (E=hυ) that L-band radio waves are emitted [1]. L-band radio waves range from 1 GHz to 2 GHz. As we have seen earlier, besides RRL, hydrogen atoms can also emit at 1.4 GHz through HFST. Further, there are a few other L-band transitions possible due to atomic hydrogen fine-structure transitions (FST). However, we restrict our interest to RRL in the L-band range. All of these L-band emissions by hydrogen atoms are summarized in Table I [1], [20].

TABLE I
L-BAND EMISSION BY HYDROGEN ATOM

| Frequency (GHz) | Transition in Energy Levels | Transition Type |
| --- | --- | --- |
| 1.077 | 3D 5/2 → 3P 3/2 | Fine-structure (Ref. [20]) |
| 1.238 | 4P 3/2 → 4S 1/2 | Fine-structure (Ref. [20]) |
| 1.368 | 4D 3/2 → 4P 1/2 | Fine-structure (Ref. [20]) |
| 1.420 | Ground state spin transition | Hyperfine-structure (Ref. [1]) |
| 1.013, 1.968 | n+1 → n (n varies from 149 to 186) | Hnα (total of 38 lines, only the first and last emissions are indicated) RRL (Ref. [1]) |

## 3  SIMULATION MODEL

The simulation model that we have developed addresses two functional parts. The first part simulates gamma water radiolysis, and the latter part simulates L-band RRL emission. Each of these models is discussed below.

### 3.1  Gamma water radiolysis

#### 3.1.1  Simulation Algorithm

The gamma radiolysis simulation framework, that we have implemented, mimics a cubical radiation chamber of side one meter, filled with unaerated water at 25 °C and exposed to an impulse of a certain number ($N_{photon}$) of 1.17 MeV (energy of a gamma photon from Co-60 radio-isotope, generated artificially in nuclear reactors) gamma photons located at the cube center. The Monte Carlo (MC) simulation of the transport of gamma photons and the electrons produced by them in liquid water and subsequent computation of atomic hydrogen population is performed using the gamma water radiolysis simulation framework. However, the LabVIEW implementation of this simulation framework is being carried out for the first time by the authors deriving inspiration from the MC particle transport simulation algorithms reported in [21]–[25]. The parameters used for simulation are listed in Table II. The LabVIEW-based code models the early "physical" and "physicochemical" stages of radiation interaction up to ~ one ps of track development in a 3D geometrical environment. The objective of the simulation is to obtain the relationship between the number of incident gamma photons and the population of excited hydrogen atoms, responsible for L-band RRL emission.

The essential features of the MC simulation method are briefly stated here. The stochastic description of any given problem like the one stated in the previous paragraph is the foremost requirement to implement MC simulation code [25]. It is possible to model the gamma water radiolysis problem as a sequence of discrete or continuous events [25]. The probability distribution function is used in the case of discrete events. On the other hand, probability density function (PDF) is utilized in case of continuous events. In either case, random sampling is applied to determine possible outcomes of the event [48]. This process of random sampling is repeated several thousand times to get an accurate description of the random event [25].

TABLE II
SIMULATION PARAMETERS FOR GAMMA WATER RADIOLYSIS

| Simulation parameter | Value |
| --- | --- |
| Radiation chamber geometry | cube |
| Radiation chamber dimension | 1 m x 1 m x 1 m |
| Maximum incident gamma photon energy | 1.17 MeV |
| Maximum number of photons | $3 \times 10^6$ |
| Gamma energy tracking cut-off limit | 100 keV |
| Electron energy tracking cut-off limit | 10 eV |
| Nature of gamma source | Pulse duration – 100 ms |
| Location of gamma source | Cube centre |
| Radiolysis stages simulated | Physical, physicochemical |
| Programming environment | Communications LabVIEW 1.1 |
| Post-processing | MATLAB (R2018b) |
| User input | LabVIEW GUI |
| User output | LabVIEW GUI, textfile – csv format |
| Look-up tables | Text file – csv format |

A basic understanding of the sampling process applied to discrete events is given here. Consider the experiment of tossing an unbiased coin. As we know, this experiment has two possible outcomes, which we can represent as – {head, tail}. Further, the probability of each of these outcomes is 0.5. A decimal number is randomly obtained from a uniform distribution ranging from zero to one. The obtained number is called a sample and is assigned to a variable called the sampling variable, $\eta$. If $\eta$ is less than or equal to 0.5, then the outcome can be treated as head. However, if $\eta$ is greater than 0.5, then the outcome can be considered as tail. To further understand the sampling process, let us consider an experiment with three possible outcomes. The experiment is a particle (electron or photon) collision event with a water molecule consisting of three possible outcomes, namely ionization, excitation, and elastic collisions. Let the probabilities of occurrence of these outcomes be 0.6, 0.3, and 0.1 respectively. Let us perform the sampling process, which will assign a value between zero and one to $\eta$. If $\eta$ is less than or equal to 0.6, then the outcome is treated as ionization collision. The outcome is considered as excitation collision if $\eta$ is between 0.6 and 0.9 (obtained by summing 0.6 and 0.3). Finally, the outcome is considered as elastic collision if $\eta$ is greater than 0.9 and less than equal to one (obtained by summing 0.6, 0.3 and 0.1). Thus, we can see that the random sampling process leads to one of the possible outcomes.

When applied to a continuous event, the random sampling process requires the PDF to be recast into an appropriate form, namely the cumulative density function (CDF). The recast is accomplished by integrating the PDF over the applicable range of corresponding variable(s) to obtain the CDF. For the example PDF given in Eq. 10 (r is a variable and μ is a constant), the required CDF is given by Eq. 11. The CDF is equated to the sampling variable $\eta$. The variable (in this example r) is written in terms of $\eta$ by inverting the CDF. This is illustrated using Eq. 12, 13. Now $\eta$ is uniformly sampled over [0,1] to obtain a possible outcome (obtaining a numerical value in a valid range for the unknown variable r from the known function pI).

$$P(r) = \mu e^{-\mu r} \tag{10}$$

$$c(r) = \int_{r'=0}^{r} p(r')dr' = 1 - e^{-\mu r} \tag{11}$$

$$1 - e^{-\mu r} = \eta \tag{12}$$

$$\Rightarrow r = \frac{-1}{\mu} \log_e(1 - \eta) \tag{13}$$

The simulation of water radiolysis is accomplished by using a transport model for the gamma photons and electrons [21]. The transport model is suited for simulation using the MC technique since it employs PDF or probability distribution of four specific events that the particles undergo along their path of travel. The four events, arranged sequentially, are free path travel, collision, energy loss, and scattering. As could be inferred from the discussion above, the random sampling process assigns a value to the physical quantities describing these four events – namely free path length, nature of collision, amount of energy loss, and angle of scattering appropriately. One iteration of the random sampling process can move the gamma photon or electron by one step. The technique of random sampling and moving the particle by another step is continued until the particle's energy drops below a threshold or crosses the boundary of the radiation chamber. The electrons collide with water molecules during each step, producing excited ($H_2O^*$) and ionized species ($H_2O^+$). With this background, we can study the simulation algorithm, which is explained below.

*3.1.1.1    Gamma photon transport*

The simulation algorithm for the transport of gamma photons is explained using the schematic presented in **Figure 2** and the flow chart depicted in **Figure 3**. The initial energy and the total number of gamma photons are obtained as user inputs. Consider a gamma photon of initial energy $E_{gi}$, traveling a free-flight length r, before interacting with a water molecule, arbitrarily located at the origin of the coordinate system (**Figure 2**). The free-flight length r is obtained by applying random sampling to the PDF pI given by Eq. 10 [22]. In Eq. 10, µ is the linear attenuation coefficient of water to gamma radiation. The initial azimuth, $\phi_{gi}$, and elevation angle, $\theta_{gi}$, of incidence for the gamma photon shown in **Figure 2** are obtained using Eq. 14 and Eq. 15 respectively [23]. In Eq. 14, 15, and throughout this section $\eta$, the sampling variable is assigned a uniformly sampled value in the interval [0,1] for each iteration. The gamma photon is transported by a distance r in the direction $(\phi_{gi}, \theta_{gi})$.

$$\phi_{gi} = 2\pi\eta \tag{14}$$

$$\theta_{gs} = \pi\eta \tag{15}$$

The gamma photon after the interaction is scattered at an azimuthal angle, $\phi_{gs}$, given by Eq. 16 and an elevation angle, $\theta_{gs}$, given by Eq. 17 [23]. Subsequent elevation scattering angle, $\theta'_{gs}$, is different from the first interaction elevation scattering angle, $\theta_{gs}$, and is computed using Eq. 18 [23].

$$\phi_{gs} = 2\pi\eta \tag{16}$$

$$\theta_{gs} = \cos^{-1}[(2\eta - 1)^2 + (1 - (2\eta - 1)^2)\cos\phi_{gs}] \tag{17}$$

$$\theta'_{gs} = \cos^{-1}\left[(2\eta - 1)\cos\theta_{gs} + \sqrt{1 - (2\eta - 1)^2}\sqrt{1 - \cos^2\theta_{gs}}\cos\phi_{gs}\right] \tag{18}$$

The energy lost, $\delta E_{gi}$, by the gamma photon during the interaction is computed by using Eq. 19 [22]. In Eq. 19, $r_e$ is the classical electron radius, $m_e$ is the electron rest mass and c is the speed of light in vacuum.

$$\delta E_{gi} = \frac{E_{gi}}{1 + \frac{m_e c^2}{E_{gi}(1 - \cos\theta_{gs})}} \tag{19}$$

Each interaction of the gamma photon with the water molecule will result in the generation of a Compton electron. Hence, an entry for the Compton electron is created, and its energy is set to $\delta E_{gi}$. Further, the Compton electron is assigned the spatial coordinate of the parent gamma photon. The Compton electron is treated as a primary electron from this point onwards. The primary electron can be defined as an incident electron with sufficient energy to excite or ionize a water molecule [21]. The primary electron record is stored in the primary electron stack (PES) for further processing. Besides, the gamma photon energy is updated by reducing the energy lost by the gamma photon. If the gamma photon crosses the radiation chamber boundary or the revised gamma photon energy is less than 100 keV, the terminating conditions for the gamma photon is considered to be satisfied and the next available gamma photon is taken up [21]. However, if this is not the case, the free-flight length, r, is obtained again by applying random sampling to the PDF given by Eq. 10. The gamma photon is transported by the distance of r in the direction ($\phi_{gs}, \theta_{gs}$). The gamma photon scattering direction, subsequent energy loss, and other events are repeatedly determined as presented in the flow chart (**Figure 3**) for transporting the gamma photon until the terminating conditions are met.

*3.1.1.2  Electron transport*

The key to the simulation of electron transport is the utilization of electron collision cross-section data. The authors have employed the cross-section data published in the literature for simulating electron collision with water molecules [19]. The readers are referred to the previously published data on the cross-section offered by water molecules to various types of electron collision as a function of incident electron energy [19]. As noted earlier, the authors refer the Compton electrons as those from the PES. The primary electron has incident energy, T, and a spatial coordinate associated with it. At the designated spatial coordinate, the primary electron undergoes a collision with a water molecule for the incident energy T. The three types of collisions that the electron undergo upon interaction with water molecule are excitation, ionization, and elastic [19]. The probability distribution function for the collision type, $P_i(T)$, is given by Eq. 20 [21].

$$P_i(T) = \frac{\sigma_i(T)}{\sum_{i=1}^{3} \sigma_i(T)} \tag{20}$$

In Eq. 20, subscript I is a variable that represents the excitation (i=1), ionization (i=2), elastic (i=3) events and $\sigma$ is the collision cross-section. The type of collision is determined by applying random sampling to the probability distribution function. The subsequent algorithm flow depending on the type of electron collision is discussed below and illustrated in the flow chart (**Figure 4**).

*3.1.1.2.1  Excitation collision*

Excitation collision can result in three types of excitation species – $\tilde{A}^1B_1$, $\tilde{B}^1A_1$, and plasmon [21]. The excitation species is determined by applying random sampling to the probability distribution function, $P_{ex-i}(T)$, given by Eq. 21 [21]. In Eq. 21, $\sigma_{ex}^i$ is the excitation cross-section. The variable I represents the species $\tilde{A}^1B_1$ (i=1), $\tilde{B}^1A_1$ (i=2), and plasmon (i=3).

$$P_{ex-i}(T) = \frac{\sigma_{ex}^i(T)}{\sum_{i=1}^{3} \sigma_{ex}^i(T)} \tag{21}$$

The energy lost by the primary electron, $W_{ex}$, is obtained by performing random sampling on the excitation energy loss differential cross-section PDF given by Eq. 22 [19], [26]. In Eq. 22, the cross-section function, $\rho(W_{ex})$ is given by Eq. 23 [19], [26] and Gaussian function, $f_i(W_{ex})$ is given by Eq. 24 – 26 [19], [26]. Further in Eq. 22, m assumes a value of two for T less than 50 eV and m is assigned a value of four for T greater than or equal to 50 eV [19], [26].

$$\frac{D\sigma_{ex}^i(T)}{dW_{ex}} = \rho(W_{ex})W_{ex}f_i(W_{ex}) \ln\left[\frac{mT}{W_{ex}}\right] \quad (22)$$

$$\rho(W_{ex}) = \frac{4\pi a_0^2}{T}\left(\frac{R}{W_{ex}}\right)^2 \quad (23)$$

$$f_i(W_{ex}) = f_{o,i}\sqrt{\frac{\alpha_i}{\pi}}\exp\left[-\alpha_i(W_{ex} - W_{0,i})^2\right] \text{ for } \tilde{A}^1B_1, \tilde{B}^1A_1 \quad (24)$$

$$f_i(W_{ex}) = f_{o,i}\alpha_{pl}\frac{e^\tau}{(1+e^\tau)^2} \text{ for plasmon} \quad (25)$$

$$\tau = \alpha_{pl}(W_{ex} - W_{0,i}) \quad (26)$$

In Eq. 23, $a_0$ is Bohr radius and R is Rydberg energy (13.6 eV). The values for the parameters $f_{o,i}$, $\alpha_i$ and $W_{0,i}$ used in Eq. 24-26 are provided in Table III [19].

TABLE III
PARAMETER VALUES FOR COMPUTING ENERGY LOST
BY PRIMARY ELECTRON DURING EXCITATION [19]

| Parameter | $\tilde{A}^1B_1$ | $\tilde{B}^1A_1$ | Plasmon |
|---|---|---|---|
| $f_{o,i}$ | 0.0187 | 0.0157 | 0.7843 |
| $\alpha_i$ | 3 (eV$^{-2}$) | 1 (eV$^{-2}$) | 0.6 (eV$^{-1}$) |
| $W_{0,i}$ | 8.4 | 10.1 | 21.3 |

The energy lost by the primary electron, during excitation is transferred to the excited water molecule. Further, the spatial coordinate of the water molecule is set to the coordinate of the primary electron. The primary electron energy is updated by reducing $W_{ex}$ and $W_{soft}$ from the primary electron energy before excitation. $W_{soft}$ is the average energy lost by the primary electron due to the numerous soft collisions occurring between two consecutive collisions between the primary electron and water molecule. Frequently the value used for $W_{soft}$ is 5 eV [21].

If the revised primary electron energy is greater than 10 eV, the free path length, r, of primary electron is determined by performing a random sampling on the PDF given by Eq. 27 [21]. Further, the azimuth scattering angle, $\phi_p$, is obtained from Eq. 28 [21]. Elevation scattering angle, $\theta_p(T)$, is determined from Eq. 29-30 [21]. In Eq. 28-29, $m_e$ is electron mass and W is equal to $W_{ex}$. The primary electron is transported by a distance of r in the direction $(\phi_p, \theta_p)$ and undergoes a collision with a water molecule at the transported point. The process of determining the collision type and choosing the appropriate course of action is already explained earlier and is also presented using the flowchart (**Figure 4**).

$$p(r) = \sigma_{ex}^i(T)e^{-\sigma_{ex}^i(T)r} \quad (27)$$

$$\phi_p = 2\pi\eta \quad (28)$$

$$\theta_p(T) = \cos^{-1}\left[\sqrt{\frac{W}{T}\frac{T+2m_ec^2}{W+2m_ec^2}}\right] \quad \text{for W/T} > 1000 \tag{29}$$

$$\theta_p(T) = \cos^{-1}\left[\sqrt{\frac{T-W}{T}\frac{T+2m_ec^2}{(T-W)+2m_ec^2}}\right] \quad \text{for } \frac{W}{T} < 1000 \tag{30}$$

If the revised primary electron energy falls below 10 eV, the primary electron is removed from the PES. Further, the next primary electron entry from the PES is considered for processing.

3.1.1.2.2    Ionization collision

Ionization collision can result in five types of ionization species – $1b_1$, $3a_1$, $1b_2$, $2a_1$, $1a_1$. The ionization species is determined by applying random sampling to the probability distribution function, $P_{ion-i}(T)$, given by Eq. 31 [21]. In Eq. 31, $\sigma_{ion}^i$ is the ionization cross-section. The variable I represents the species $1b_1$ (i=1), $3a_1$ (i=2), $1b_2$ (i=3), $2a_1$ (i=4) and $1a_1$ (i=5).

$$P_{ion-i}(T) = \frac{\sigma_{ion}^i(T)}{\sum_{i=1}^{5}\sigma_{ion}^i(T)} \tag{31}$$

The energy lost by the primary electron, $W_{ion}$, is obtained by performing a random sampling on the ionization energy loss differential cross-section PDF given by Eq. 32 [19]. In Eq. 32, $N_i$ is the number of electrons in the $i^{th}$ molecular orbital of water, $a_o$ is Bohr radius and R is Rydberg energy.

$$\frac{D\sigma_{ion}^i}{dW_{ion}} = \frac{S_i}{I_i}(F_1(t)[T_1^3 + T_2^3 - T_3^3] + F_2(t)[T_1^2 + T_2^2 - T_3^2]) \tag{32}$$

Where $T_1 = \dfrac{1}{1+w}$ $\qquad F_1(t) = \dfrac{A_1 \ln(t)}{t+B_1}$ $\qquad w = \dfrac{W_{ion}}{I_i}$

$T_2 = \dfrac{1}{t-w}$ $\qquad F_2(t) = \dfrac{A_2}{t+B_2}$ $\qquad A_1 = 1.31; A_2 = 0.37; B_1 = B_2 = 0$

$T_3 = \dfrac{1}{\sqrt{(1+w)(t-w)}}$ $\qquad t = \dfrac{T}{I_i}$ $\qquad S_i = 4\pi a_0^2 N_i\left(\dfrac{R}{I_i}\right)^2$

The ionization collision results in an electron getting knocked off from the water molecule called a secondary electron. Therefore, an entry is created for the secondary electron to track it. The spatial coordinate of the secondary electron is set to the coordinate of the primary electron. Further, the secondary electron is assigned the energy $W_{ion}$ reduced by the sum of $W_{soft}$ and binding energy. The binding energies for $1b_1$, $3a_1$, $1b_2$, $2a_1$, $1a_1$ are 12.62 eV, 14.75 eV, 18.51 eV, 32.4 eV, 539.7 eV respectively [19], [27]. If secondary electron energy is greater than 10 eV, the secondary electron entry is added to the PES. Further, the primary electron energy is updated by reducing $W_{ion}$ and $W_{soft}$ from the primary electron energy before ionization.

If the revised primary electron energy falls below 10 eV, the primary electron is removed from the PES. Further, the next primary electron entry from the PES is considered for processing. However, if the revised primary electron energy is greater than 10 eV, the free path length, r, of primary electron is determined by performing a random sampling on the PDF given by Eq. 33 [21]. Further, the azimuth scattering angle, $\phi_p$, is obtained from Eq. 28 [21]. Elevation scattering angle, $\theta_p(T)$, is determined from Eq. 29-30 [21]. In Eq. 29-30, W is equal to $W_{ion}$. The primary electron is transported by a distance of r in the direction $(\phi_p, \theta_p)$ and undergoes a collision with a water molecule at the transported point.

$$p(r) = \sigma_{ion}^i(T)e^{-\sigma_{ion}^i(T)r} \tag{33}$$

3.1.1.2.3    Elastic collision

If the electron collision type with water molecule is elastic, the free path length, r, of primary electron is determined by performing a random sampling on the PDF given by Eq. 34 [21]. Further, the azimuth scattering angle, $\phi_p$, is obtained from Eq. 28 [21]. Elevation scattering angle, $\theta_p(T)$, is determined from Eq. 29-30 [21]. In Eq. 29-30, W is equal to zero, since there no energy is lost during an elastic collision. The primary electron is transported by a distance of r in the direction $(\phi_p, \theta_p)$ and undergoes a collision with a water molecule at the transported point.

$$p(r) = \sigma_{el}(T)e^{-\sigma_{el}(T)r} \tag{34}$$

TABLE IV
BRANCHING RATIOS OF EXCITED SPECIES [11], [16]

| Equation | $\tilde{A}^1B_1$ | $\tilde{B}^1A_1$ | Plasmon |
|---|---|---|---|
| $H_2O^* \to H_2O$ (4) | 35% | 35 % | |
| $H_2O^* \to H + OH$ (5) | 65 % | 43.22 % | 8 % |
| $H_2O^* \to H_2 + O$ (6) | | 10.08 % | |
| $H_2O^* \to 2H + O$ (7) | | 11.7 % | |
| $H_2O^* \to H_2O^+ + e^-$ (8) | | | 92 % |
| **Total** | 100 % | 100 % | 100 % |

*3.1.1.3    Total population of H atoms*

As already mentioned, the excited species ($H_2O^*$), indicated in Eq. 3, further disintegrate as per the reactions specified by Eq. 4 – Eq. 8. The branching ratios given in Table IV determine the percentage of species that undergo each of these five channels of dissociation [11], [16]. Utilizing the information in Table IV, we can infer that the total number of hydrogen atoms, $P_H$ at 1 ps, is given by Eq. 35. Since we have already tagged the excited water molecules by their species type ($\tilde{A}^1B_1$, $\tilde{B}^1A_1$, or plasmon), we can apply Eq. 35 to find the atomic hydrogen population at 1 ps. In Eq. 35, P denotes the population of the species designated by the subscript.

$$P_H = 0.65 P_{\tilde{A}^1B_1} + 0.6662 P_{\tilde{B}^1A_1} + 0.08 P_{plasmon} \tag{35}$$

*3.1.2    Outline of LabVIEW code*

A bird's eye view of the LabVIEW code developed by the authors utilizing the simulation algorithm is shown in **Figure 5**. The upper-level code uses about 115 subroutine files to simulate and validate gamma water radiolysis. As is evident from **Figure 5**, there are three major blocks – processing of gamma photons, primary electrons, and computation of total atomic hydrogen population. The processing of gamma photon block obtains free path length (GP1) and scattering angle (GP2) for each gamma photon. Further, it moves the photon to the next step (GP3) and a collision is registered. The energy loss during the collision is sampled (GP4). The Compton electrons produced during the interaction are simulated (GP6) and pushed to the PES (GP7). The processing is carried out for the number of gamma photons (tracked by GP5 and the for loop) input by the user.

The processing of primary electrons block picks up each simulated primary electron from the stack (PE1). The program control consults the cross-section data created as look-up tables (PE2). The sampling of collision type is performed utilizing the cross-section data (PE3). Depending on the type of collisions – excitation, ionization, and elastic; sampling of free path length (PE4), energy loss (PE4), and scattering angle (PE5) is made for the simulated primary electron. The primary electron is moved to the next step (PE6) using the sampled values. The secondary electrons created during ionization are simulated and pushed to the PES (PE7). The processing of primary electrons is repeated until the PES is empty (PE8). Finally, the total atomic hydrogen population is computed using Eq. 35. Besides, there are several other information such as the energy and spatial location of the primary electrons, excited and ionized water molecules, the spatial location of hydrogen atoms, which are stored in csv (comma-separated values) file format, suitable for post-processing by a program such as MATLAB.

### 3.1.3 *Computation of L-band RRL H atom population*

The L-band RRL H atom population is a subset of the total H atom population. Hence, we need to find the fraction of the total H atom population contributing to the L-band RRL emission. However, the H atom population is a function of time elapsed after the gamma photon's interaction with a water molecule (as is evident from **Figure 1**). Therefore, we need to define a more precise time window of interest during which the H atom population must be determined. A significant L-band RRL emission occurs only after a few hundred microseconds following the gamma photon interaction with water molecules, which can be deduced from the mean-life time of RRL emission, $\tau_n$ given by Eq. 36 [12]. Specifically, between 0.1 $\tau_n$ to $\tau_n$, about 60 % of RRL emission occurs. But $\tau_n$ varies from about 1 ms to 3 ms for LB-RRL levels (n+1 to n, with n varying from 149 to 186). Therefore, the crucial time window (ctw) of the LB-RRL emission is 0.1 ms to 3 ms after the gamma photon interaction.

$$\tau_n = 4 * 10^{-10} n^3 \ (s) \tag{36}$$

It is reported that at least fifty percent of the excited water molecules with energy greater than 18.709 eV (dissociation limit principle for n=10) will dissociate to a hydrogen atom at n greater than or equal to 10 [2], [7], [28]. Further, these experiments have revealed that the radiolysis of water leads to a population inversion of the excited hydrogen atoms. The doubling of the population occurs when the principal quantum energy level drops by about two eV between the appropriate energy levels [6], [29], [30]. However, the energy difference between n=10 and n=300 is merely 0.13 eV. Hence, we can regard that the level between n=10 and n=300 is equally populated as a first approximation. Besides, in the simulation of water radiolysis, we track the energy of excited water molecule species. Hence, we can find the number of hydrogen atoms produced during radiolysis for each of the quantum levels from n=10 to n=$n_{max}$, which includes the population of n=149 to n=186 ($P_{LB-RRL}$) for our analysis using Eq. 37. In Eq. 37, $n_{min}$=10, and $n_{max}$=300, $G_H$ is the yield of hydrogen atoms, P (n in the range of $n_{min}$ to $n_{max}$) is the total atomic hydrogen population for levels $n_{min}$ to $n_{max}$ at 1 ps after the gamma photon interaction. Here $n_{max}$ is the upper limit of hydrogen principal quantum level. Radio-astronomers have observed atomic hydrogen RRLs with excitation as high as $n_{max}$ =300 and emissions beyond $n_{max}$ = 300 have been rarely observed [1]. Hence, our analysis limits $n_{max}$ to 300. It should be noted that the yield information is available only up to 1 ms in the literature (see **Figure 1**), but the crucial time window of LB-RRL emission extends beyond 1 ms up to 3 ms. Fortunately, H˙ atomic population remain intact even beyond 1 ms up to few tens of milliseconds and likely to partake in the L-band RF emission process [27], [31]. Therefore, when computing the population for more than 1 ms, the yield values at 1 ms were used.

$$P_{LB-RRL} = \frac{P(n=n_{min}:n_{max})\frac{G_H(t=ctw)}{G_H(t=1\text{ ps})}}{n_{max}-n_{min}} \qquad (37)$$

*3.2    L-band RRL emission*

This section will first analyse the feasibility of L-band RF emission due to RRL during the gamma radiolysis of water, then compute L-band RF generated power, examine radio spectral index, and RRL broadening.

*3.2.1    Feasibility of RRL emission*

We want to investigate the possibility of ionization of the Rydberg hydrogen atoms before they emit L-band RF photons. The radius of the Rydberg atoms of n=149 to 186 range exceeds one micro-meter [1]. The lifetime of such massive particles necessitates discussion. Two specific cases have emerged from the collisional studies of Rydberg atoms that have been actively carried out since the 1980s [32]–[34]. These cases are (1) the translational velocity ($V_t$) of the Rydberg atom is either very small or comparable to the electron orbital velocity ($V_o$) (2) $V_t$ is much greater than $V_o$ [32]–[34].

In the first case, the atomic core and the electron interact with colliding particles independently. Further, the inelastic collision cross-section of the atomic core is so small that it can be neglected. Hence, essentially, in this case, the inelastic collision cross-section of the Rydberg atom, despite it being huge, reduces to that of a free electron [32]–[34]. The simulation conducted by us reveals that the Rydberg hydrogen atoms have kinetic energy ranging from zero eV/u to about 30 eV/u with a peak at zero eV/u. Further, the kinetic energy distribution of Rydberg hydrogen atoms created during the radiolysis of water, under the impact of 10 keV electrons, ranges from zero eV/u to about 100 eV/u with a peak at seven eV/u [5], [35]. These energies correspond to a $V_t$ that is of the same order of magnitude as $V_o$. For $V_t$ to be much greater than $V_o$, the kinetic energy of the hydrogen atom should exceed 2.5 keV/u [32]. Thus, it can be concluded that the radiolysis of water exposed to gamma energy less than or equal to 1.17 MeV belong to the first category.

In the second case, the core is no longer a spectator, and both the electron and the core determine the collisional properties [32]–[34]. Besides an increased collision cross-section of the core due to enhanced velocity, core collisions transfer energy to electrons, causing depopulation of the Rydberg state [32]–[34].

$$\sigma_{\text{ion}} = 5.5\pi n^4 q a_0^2 \tag{38}$$

The ionization of the Rydberg hydrogen atom is caused either due to the inelastic collision of the Rydberg hydrogen atom with another ion/electron or neutral water molecules. We will first consider the case of the ionization of the Rydberg hydrogen atom due to collision with an ion/electron. The expression for the inelastic collision cross-section of the Rydberg hydrogen atom with an ion is given by Eq. 38 [36]. This equation is valid when the translational velocity of a Rydberg atom has the same orders of magnitude as that of the orbital electron velocity (first case), which holds good for the gamma radiolysis of water discussed in this article. In Eq. 38, q is the electron charge, n is the principal quantum number and $a_0$ is the Bohr radius. Employing Eq. 38, the computed inelastic collision cross-section ranges from $3.4 \times 10^{-26}$ to $9.2 \times 10^{-26}$ cm². However, the ion density, $N_i$, for a maximum allowed separation between adjacent RRLs of 25 MHz computed using Eq. 39 is $1.6 \times 10^6$ /cm³ [37].

$$\Delta v_L^i = \frac{3}{\sqrt{\pi}} \left(\frac{h}{2\pi m}\right)^2 \sqrt{\frac{M}{kT_e}} N_i n^2 \left[\frac{3}{2} + \frac{2}{e^2} \log_e \left(\frac{2\pi}{3}\right)\right] \left[\frac{1}{2} + \log_e \left(\frac{\lambda}{n}\frac{m}{h}\sqrt{\frac{kT_e}{M}}\right)\right] \tag{39}$$

In Eq. 39, M is the mass of ion ($H_2O^+$), $N_i$ is the ion density, e is the base of natural logarithm, $\lambda$ is the emission wavelength, n is the principal quantum number, k is Boltzmann's constant, $\Delta v_L^i$ is the increase in velocity of ion and m is the mass of the electron. Further at the computed ion/electron density, the average time, $t_{\text{ion}}$, between the successive inelastic collision of the Rydberg atom with an ion can be computed using Eq. 40 and is found to be in the order of several hours [21].

$$t_{\text{ion}} = \frac{1}{v_{\text{ion}} N_i \sigma_{\text{ion}}} \tag{40}$$

In Eq. 40, $v_{\text{ion}}$ is the average relative velocity of ion with respect to the Rydberg atom. Hence, we can conclude that the probability of de-populating a Rydberg hydrogen atom by another electron or ion is very low. We now consider the possibility of ionization of Rydberg hydrogen atom by a neutral water molecule. The inelastic collision cross-section, $\sigma_w$, of the Rydberg atom (calculated for the first case for n>140) with a neutral water molecule is $10^{-32}$ cm$^2$ [38]. The average interaction interval ($t_w$) for an inelastic interaction of a Rydberg hydrogen atom with neutral water molecule, computed using Eq. 41, is several seconds [21]. Thus, the interaction time between the successive collision of Rydberg hydrogen atom with either an ion or water molecule is significantly higher than the mean lifetime of emission, which ranges from 1.3 ms to 2.6 ms (computed using Eq. 36). In this context, Rydberg hydrogen atoms, subjected to gamma radiation less than or equal to 1.17 MeV, will invariably emit L-band RF photons before getting ionized. In Eq. 41, $v_w$ is the average relative velocity of water molecules with respect to the Rydberg atom and $N_w$ is the number of water molecules present in a cubic meter volume of water.

$$t_w = \frac{1}{v_w N_w \sigma_w} \tag{41}$$

*3.2.2 L-band RF generated power*

Using the population of hydrogen atoms in the LB-RRL levels, we determine the RF-generated power I the electric field intensity of the radio wave generated due to spontaneous decay of the LB-RRL levels. The photon emission wave equivalent is obtained by modeling the spontaneous decay process as a decaying simple harmonic oscillation (SHO) of the transitioning electron at the rate equal to the emitted photon frequency [3]. The far electric field has only a non-zero component perpendicular to wave propagation's direction and is given by Eq. 42 [3] [39]. In Eq. 42, q is charge of an electron, f is wave frequency, n is principal quantum number, $a_0$ is Bohr radius, $\epsilon_1$ is the permittivity of the medium, r is the distance at which electric field has to be determined, k is the phase shift constant, $\theta$ is the angle between the direction of propagation and the axis of the oscillating charge and $\gamma_0$ is the transition probability.

$$E(r,t) = \frac{-q(2\pi f)^2 a_0 n}{4\pi \epsilon_1 r c^2} e^{-\gamma_0 t/2} e^{j(kr-(2\pi f)t)} \sin\theta \tag{42}$$

Each of the elements in the set $\varphi_{Hj}$ (all hydrogen atoms occupying the principal quantum level j, $149 \leq j \leq 186$) can now be considered as a simple harmonic oscillator. The simple harmonic oscillator can be oriented in various directions. The values for the orientation of the axis of the oscillator is sampled from a uniform distribution. We divide the surface (boundary walls) of the radiation chamber into several segments. The segment's size should be small enough so that the electric field is almost constant in that piece. The electric field, |E$_S$|, at a segment can be computed by considering the electric field from each set of $\varphi_{Hj}$ elements using Eq. 42 and summing up all the elements' contributions. The power density in each segment can be found using Eq. 43, following which the power can be found using Eq. 41. In Eq. 43, η is the intrinsic impedance of water. In Eq. 44 $A_i$ is the area of the $i^{th}$ segment. The total generated RF power is the sum of the power associated with each segment and is given by Eq. 45. In Eq. 45, N is the total number of segments.

$$P_D = \frac{|E_S|^2}{2\eta} \tag{43}$$

$$P_{Si} = P_D A_i \tag{44}$$

$$P_G = \sum_{i=1}^{N} P_{Si} \tag{45}$$

*3.2.3  Radio Spectral Index*

The radio spectral index is employed for understanding the origin of the radio emission [40]. A log-log plot of generated power to LB-RRL level frequencies, called spectral energy distribution (SED), needs to be created to obtain radio spectral index. The spectral index is the SED slope in a log-log plane. It has been established that the spontaneous H radio recombination lines have a positive value of radio spectral index while a negative value indicates stimulated emission [37], [39]–[42].

*3.2.4  RRL Broadening*

The two significant phenomena that contribute to RRL broadening are Doppler broadening and Stark broadening. The Doppler broadening is attributed to the velocity distribution gained by the turbulent emitting hydrogen atoms. The frequency broadening, f', about the center frequency $f_0$ can be computed from the velocity profile, $\upsilon$, of H atoms using Eq. 46 [41].

$$f' = f_0 \sqrt{\frac{1+\frac{\upsilon}{c}}{1-\frac{\upsilon}{c}}} \tag{46}$$

Stark broadening is also loosely referred to as pressure broadening or impact broadening [1]. The Stark broadening is produced due to the cumulative phase distortion of the emitted wave that results from electron or ion collisions with the emitting atom [1]. Stark broadening due to ion-impact collision is given by Eq. 39 [37]. Stark broadening due to electron-impact collision is given by Eq. 47 [37]. In Eq. 47, $N_e$ is the electron density in /cm$^3$ and $T_e$ is the temperature in Kelvin.

$$\Delta v_L^e = \frac{5.15^{-6} N_e n^4}{\sqrt{T_e}} \log_e(8.25. 10^{-6} T_e n) \tag{47}$$

*3.3  Gamma radiation and RF intensity – relationship*

Employing the outcome of modeling gamma water radiolysis and L-band RRL emission, we can now find the relationship between gamma radiation absorbed and the intensity of radio waves emitted due to the radiolysis process. We utilize the absorbed dose rate (ADR) to characterize the gamma radiation, while generated power is used to describe the L-band RF emission. ADR is defined as the gamma energy absorbed by medium (water) per unit mass per unit time [4], [13], [17]. ADR is computed using the gamma energy deposition map obtained from the LabVIEW MC simulation. Further, we utilize the broadening calculations to obtain the saturation region of the RF-generated power. Characterizing dependence of radio intensity with the gamma radiation is the last step in the RF source modeling of gamma radiation interaction due to water radiolysis.

## 4 RESULTS AND DISCUSSION

### 4.1 Gamma water radiolysis

#### 4.1.1 Computation of L-band RRL H atom population

The production of hydrogen atoms excited to the principal quantum levels of n = 149 to n=186 required for L-band RRL emission during water radiolysis has not been experimentally verified. However, Rydberg hydrogen atoms with excited principal quantum levels up to 300, though produced through mechanisms unrelated to water radiolysis, have been observed in deep space [1]. Besides, experimental reports confirm the production of Rydberg hydrogen atoms excited to a principal quantum level greater than ten during water radiolysis [5], [35]. However, the scope of our work is restricted to the use of simulation for computation of Rydberg L-band RRL H atom population.

The total atomic hydrogen population and the atomic hydrogen population occupying principal quantum level n=10 to n=300, at 1 ps after the gamma photon interaction as a function of the number of gamma photons obtained using LabVIEW particle simulation (Eq. 10 to Eq. 30, Eq. 35) is shown in **Figure 6.** Further, the lower limit of the number of gamma photons is set to 100 in the simulation. Below the lower limit, since the simulation process is stochastic, reproducibility of data obtained through simulation is questionable, possibly due to inadequate conformance to the law of large numbers required by probability theory. The number of hydrogen atoms created is proportional to the number of incident gamma photons. For each 1.17 MeV gamma photon, a total of about 330 hydrogen atoms and 240 hydrogen atoms for n $\geq$ 10 is created at 1 ps. Employing this result in Eq. 37, the atomic hydrogen population of each of the L-band RRL at 0.1 ms is obtained and shown in **Figure 6**. On average, one 1.17 MeV gamma photon interacting with liquid water produces about three hydrogen atoms at each L-band level. Therefore, the total number of hydrogen atoms excited by the gamma photon to all possible LB-RRL levels (n=149 to n=186) is about 120. A significant fraction of these hydrogen atoms will make an Hn-α spontaneous transition, producing L-band RF photons.

*4.1.2    Relative intensity of L-band RF emissions*

Utilizing the population of L-band emitting hydrogen atoms and transition probability, we can determine the relative intensity of L-band RF emissions using Eq. 48 [43]. In Eq. 48, $N_u$ is the population of atoms in the upper level, $\psi_{u:l}$ is transition probability from upper-level u to lower-level l. The dynamic range of the relative population of principal quantum levels varies only marginally for LB-RRL levels. Hence, it can be shown, using Eq. 48, that to the first order of approximation, the RF intensity of an emission line during water radiolysis is dependent only on the transition probability. The normalized RF intensity is computed and presented in **Figure 7**.

$$\frac{dN_{u:l}}{dt} = -N_u \psi_{u:l} \tag{48}$$

Among the various atomic hydrogen L-band emission lines produced by water radiolysis, the one with the weakest intensity is the 1420 MHz HFST emission line. FST emission lines are about a hundred times more intense than the HFST emission lines. Emissions at 1077 MHz, 1081 MHz, 1238 MHz and 1368 MHz correspond to the FST emissions. The RRL have the highest intensity, about ten orders of magnitude stronger than the spontaneous HFST line. We have chosen to analyze the RRL lines due to their large emission strength compared to the other L-band lines, and hence they have a higher detection chance. Finally, we note that the RRL intensity increases with a rise in emission frequency, which could be attributed to increased transition probability.

*4.2    L-band RRL emission*
*4.2.1    L-band RF generated power*

The RF-generated power corresponding to the interaction of the impulse of a certain number of gamma photons ($N_{photons}$) is obtained using the simulation model explained in section 3.2.2 (Eq. 43 – Eq. 45) and is presented in **Figure 8.** Precisely, the RF-generated power is calculated for one of the 38 possible L-band RRL lines, namely, $H167\alpha$ corresponding to a frequency of 1.4 GHz. High precision, low-cost antenna, detection systems, and designs are available at 1.4 GHz [44]. Further, diverse research teams have investigated the detection sensitivity and RFI at 1.4 GHz, which could be reused in our future work by choosing $H167\alpha$ [44].

Due to the random iterative nature of the simulation model, the simulation results do not show a precise linear dependence of RF-generated power on the number of gamma photons. However, there is a linear trend, and a line of best fit was obtained with an $R^2$ value of 0.85277. From the slope of the best fit line, we can infer the generated RF power corresponding to an impulse of one gamma photon to be $10^{-21}$ W. Though, at first instance, this value appears too tiny, radio telescopes routinely detect extremely feeble radio frequency signals of the order of a mJy (milli-Jansky) [45]. One milli-Jansky is $10^{-29}$ W/m$^2$/Hz, corresponding to L-band detection of an RF signal of strength $10^{-20}$ W incident on a one square meter aperture antenna.

The generated RF power and the detected RF power are different; both need to be related in this context. Further, not all the generated RF power is transmitted. The transmitted RF power will be about an order lower than the generated RF power because water has a higher relative permittivity than air. The higher relative permittivity value will result in total internal reflection, with only about 10 % of generated power transmitted [46]. Further, the transmitted RF power will undergo inverse square law attenuation [46]. At a distance of one meter away from the boundary of the radiation chamber, the received RF power would have diminished by about $4\pi$, assuming the receiving antenna has one square meter aperture and unity gain [46]. Hence, the received RF power would be two orders of magnitude lower than the generated RF power to the first degree of approximation in this context. Therefore, the generated RF power should be at least $10^{-18}$ W (two orders of magnitude higher than $10^{-20}$ W) for a successful detection. As is evident from **Figure 8**, an RF power of $10^{-18}$ W corresponds to about an impulse of 1000 gamma photons. Until now, we did not consider radio background noise. It is hard to find any region with human habitation free from radio frequency interference (RFI) [47]. The typical minimum RFI noise value is -110 dBm ($10^{-14}$ W) for the frequency of interest, namely, 1.4 GHz [47]. Hence a radio receiver, subjected to background RFI, could successfully detect an RF-generated power only greater than $10^{-12}$ W (two orders of magnitude greater than $10^{-14}$ W), corresponding to an impulse of $10^9$ gamma photons.

*4.2.2 Radio spectral index*

The generated RF power is computed for the 38 different L-band RRL lines using the scheme outlined in the previous section (L-band RF generated power) at a fixed absorbed dose rate corresponding to ten gamma photons, each of energy 1.17 MeV. The SED curve is obtained by plotting log (generated RF power per unit area per unit frequency) to log(frequency) and is presented in **Figure 9**. Since the simulation is performed using the Monte Carlo technique, a random iterative procedure, as expected, the simulated points in the graph do not fit into a perfectly straight line. Hence, a line of best fit (with $R^2$ of 0.88585) is drawn for the 38 points in the SED curve. The slope of the line gives the radio spectral index, which in this case is 1.7380. We obtained a positive value of radio spectral index, confirming that the L-band RF emission mechanism in gamma radiolysis of water is due to spontaneous radio recombination. The other emission mechanisms include stimulated radio recombination, bremsstrahlung, synchrotron emission and diffuse synchrotron.

This result is crucial since the simulation model is validated for the correct emission mechanism. Unlike spontaneous emission, stimulated emission can trigger amplification resulting in RRL masers. Though L-band radio recombination lines can undergo a "partial maser effect", they can never become true RRL masers [1]. RRL masers can occur only for transitions resulting in emission frequencies above 7.5 GHz [1]. Hence L-band RRL is always spontaneous and is associated with a positive spectral index.

The radio spectral index is vital for distinguishing spontaneous RRL from other emission mechanisms and verifying if maser action is initiated while looking for RRL lines of frequencies above 7.5 GHz. It has to be emphasized here that we can expect a significantly higher RF-generated power per interacting gamma photon at frequencies above 7.5 GHz, provided a maser action is triggered.

*4.2.3   RRL broadening*

*4.2.3.1   Maximum allowed broadening limit*

The generally adopted criterion for an emission line to be observable is its width should not exceed 30 % of the separation between adjacent RRLs [1]. The separation between adjacent RRLs is 25 MHz. Therefore, the maximum allowed broadening limit is 15 MHz (2 x 7.5 MHz).

*4.2.3.2   Doppler broadening*

The velocity profile of hydrogen atoms produced due to various incident electron energies is obtained from the energy distribution of hydrogen atoms (Gamma water radiolysis (section 3.1)) and shown in **Figure 10**. The FWHM velocity is 23 km/s and is almost independent of the incident electron energy. The precise FWHM velocities are 23.44 km/s, 23.52 km/s, 23.56 km/s, 23.59 km/s and 23.59 km/s for incident energies of 500 eV, 5 keV, 50 keV, 500 keV and 1 MeV. With an increase in incident electron energy, the number density of hydrogen atoms decreases, attributed to the decreased cross-section at higher incident energies.

The Doppler frequency broadening, computed using the FWHM velocity of 23 km/s and letting the frequency range from 1 GHz to 2 GHz in Eq. 43 is 300 kHz to 600 kHz. Therefore, the maximum Doppler broadening is 600 kHz, which is well within the acceptable broadening limit. Hence it is evident that Doppler broadening does not hamper L-band RF detection for gamma energies at least up to 1.17 MeV.

*4.2.3.3   Stark broadening*

The stark broadening can be computed using Eq. 44 (ion-broadening equation). As can be seen from the equation, the broadening increases with an increase in ionic concentration. During gamma radiolysis of water, ion contribution is due to hydrated electrons. The maximum allowed broadening of 15 MHz is reached when the hydrated electron concentration reaches $1.6 \times 10^6$ per $cm^3$. However, the hydrated electron density is proportional to the absorbed dose rate. Thus, we can conclude that the stark broadening places an upper limit on the absorbed dose rate. When the absorbed dose rate is increased to the extent that the hydrated electron density reaches $1.6 \times 10^6$ per $cm^3$, broadening of the RRL lines will result in substantial overlap between adjacent lines, resulting in unreliable detection.

*4.3  Gamma radiation and RF intensity – relationship*

The hypothetical cube of water was exposed to an impulse of $N_{photons}$ at t=0. The radiation transport simulation was carried up to 1 ps. The absorbed dose is computed by dividing the total energy lost by $N_{photons}$ by the mass of water in the radiation chamber. Further, all the primary species produced by low LET radiation, such as gamma radiation, will ultimately recombine in pure deaerated water. Hence, the water relaxes entirely to its initial state after about 100 ms [13]. Therefore, the absorbed dose rate was obtained by dividing the absorbed dose due to $N_{photons}$ with a relaxation time of 100 ms.

The variation of $H167\alpha$ RF-generated power with absorbed dose is presented in **Figure 11**. The hardware resources on a core i7 personal computer with 16 GB RAM could support simulation up to a dose rate of 0.2 µGy/h. Hence to predict the system's behavior to a range of 100 Gy/h, a curve was fit to the simulation data using a polynomial function of degree one, which has the least root mean square error value over the simulated measurement range $1.96 \times 10^{-19}$ W to $1.09 \times 10^{-14}$ W. The curve fit has an $R^2$ value of 0.85364. Further, the equations of the curve for the three different regions (A – linear, B – transition and C – saturation) are shown in the figure. As is evident from the figure, the RF-generated power has linear dependence on the absorbed dose.

Due to stark broadening, at an absorbed dose of about 1.13 Gy/h, the RF power generation rate drops. The stark broadening effect masks about 63 % of RF emissions. Further, at an absorbed dose of about 8.93 Gy/h, RF power generation saturates to a steady value indicating that the stark broadening masks the entire RF emission. Beyond about 9 Gy/h, water at 25 ℃ cannot be employed for dose rate measurements. However, 9 Gy/h is a relatively large dose rate value, and hence for numerous useful measures, a dose rate of 9 Gy/h is sufficient.

To appreciate that measuring up to 1 Gy/h is sufficient for most practical cases, let us analyze the following dose rate scenarios. The dose rate produced by background radiation ranges from 0.005 to 0.3 µGy/h [48]. But, steady dose rates below 0.1 uGy/h are challenging to measure, even using a conventional detector. While regions with dose rates greater than ten mGy/h are classified as exclusion areas in nuclear facilities, those with dose rate exposures higher than 100 mGy/h are considered high-dose regions [48]. Accidental exposure of the nuclear facility personnel to such high-dose regions, with beam turned-on, could result in serious health issues [49]. While the highest dosage recorded, 32 km away and 3 hours after the Trinity bomb's fallout, was about 190 mGy/h, the radiation level inside a BWR reactor's containment vessel (Fukushima power plant), six years after the meltdown, was more than 10 Gy/h [48].

We had seen that for a one-meter squared unity gain receiver antenna located one meter away from the radiation chamber, the minimum detectable generated power without RFI and with RFI is about $10^{-18}$ W and $10^{-12}$ W, respectively. From **Figure 11**, we can note that the absorbed dose rates for these values of generated power are about 0.1 nGy/h and 0.1 mGy/h, respectively. It should be noted that the minimum absorbed dose rate detectable by a conventional gamma radiation detector is 0.1 µGy/h. Hence a custom-built RF-based detector in a specialized RFI-shielded environment could outperform a traditional detector. Nevertheless, a cost-effective RF detector of nominal dimensions (precisely 1 m$^2$ aperture area) operating at ambient conditions with RFI not exceeding -110 dBm can detect only dose rates above 0.1 mGy/h when located at a reasonable distance of one meter away from the radiation chamber. The discussion of the RF-based detector is intended to provide a fair account of RF-based detection sensitivity compared to a traditional detector. However, an accurate assessment of detection sensitivity is primarily dictated by the instrumentation capabilities of the RF receiver, which the authors intend to carry out in the future and is beyond the scope of this article.

Quantifying the RRL emission as discussed in this article will pave the way for the development an RF-based gamma radiation detector. Such a detector can offer certain distinct advantages over the conventional gamma radiation detector. One of the disadvantages of the conventional gamma radiation detector is that it should be located within a few tens of meters from the source for reliable detection, because gamma radiation undergoes both inverse square law decay and exponential attenuation, resulting in a smaller field of view [50]–[52]. However, the RF signal strength undergoes dominantly only inverse square law decay in free space facilitating the radio waves to travel even further than the gamma radiation [46]. Furthermore, significant advances have been achieved in the detection ability of radio telescope in the quest for radio-imaging distant astronomical bodies [52]. In the order $10^{-20}$ watts and more diminutive, a weak radio-frequency signal could be detected by radio telescopes by utilizing large antenna arrays and the principle of radio interferometry [1]. Besides the possibility of remote radiation monitoring, we were also motivated to take up this study because water is an abundant resource on our planet. About three-fourth of the earth's surface is covered by water. Water is present in the air in the form of water vapor. Besides, many nuclear power plants (NPP) are located near water bodies. The water body around NPP could act as a "natural" active material for gamma interaction both in liquid and vapor forms facilitating gamma radiation detection using RF methods. RRL emission has been observed and studied, hitherto only at astronomical distances. Our studies could open up the possibility of monitoring RRL emissions at locations close to the earth's surface.

5   **CONCLUSION**

This work reports a Monte Carlo simulation study on the gamma radiolysis of water performed via LabView implementation of the process. The hydrogen atoms produced during gamma radiolysis of water emit L-band RRL also among other possible transitions. These observations reported in literature have been taken into account in this simulation study focused towards L-band emission. The population and spatial distribution of hydrogen atoms produced during gamma photon interaction with water was computationally determined here. Specifically, the simulation model enabled to distinguish L-band RRL emitting hydrogen atoms (Rydberg hydrogen atoms) from the other atomic hydrogen population. A detailed analysis was presented to show that the Rydberg hydrogen atoms produced during gamma radiolysis of water do not ionize and will undergo transitions from $(n+1)^{th}$ level to $n^{th}$ level (where n is the principal quantum number in the range of 149 to 186), leading to RRL emission. Furthermore, the quantitative analysis enabled computation of L-band RF power. The radio spectral index was calculated to validate the modeled RRL emission mechanism. Finally, RRL broadening and the method adopted to relate the intensity of radio waves emitted during the gamma radiolysis of water to the absorbed gamma dose rate were discussed. Specific results obtained from this simulation study are summarized below.

On an average, one 1.17 MeV gamma photon interacting with 1 m$^3$ liquid water produces about 120 L-band RF emitting Rydberg hydrogen atoms. The generated 1.4 GHz L-band RF power corresponding to an impulse of $10^9$ gamma photons is $10^{-12}$ W. The generated L-band RF power is proportional to 1.17 MeV gamma photons and has a slope of $10^{-21}$ W per gamma photon. Three distinct regions were identified in the generated RF power-absorbed gamma dose rate characteristics. These regions are the linear region extending up to 1.13 Gy/h, the transition region between 1.13 Gy/h and 8.93 Gy/h, and the saturation region beyond 8.93 Gy/h. The transition and the saturation region are a result of Stark broadening. The computed radio spectral index is 1.738, and it is inferred that the nature of L-band RRL emission mechanism is spontaneous decay.

Thus, our work presents a theoretical source model for radio wave emission during gamma radiolysis of water. The authors believe this work will help evolve experiments that can demonstrate L-band RF emission during gamma radiolysis of water. Further, the RF emission and subsequent detection during gamma radiolysis of water could open up numerous avenues in advancing radiation monitoring. An RF-based detector can likely be placed at further stand-off distances (increased altitude measurements) than the conventional gamma radiation detector due to radio waves' nature and also due to the tremendous progress achieved in radio imaging instrumentation frequently employed as radio-telescopes. Developing and deploying the appropriate technologies at increased altitudes will enhance the field of view, which, in turn, will reduce the need for deploying several air-borne monitoring vehicles at lower altitudes. Further, the personnel operating farther away from the accident site will be shielded from exposure to harmful gamma emissions in the vicinity of accident. Moreover, this paves the way for the air-borne surveys to be carried out perpetually (for example, using a satellite constellation), which hitherto, had been carried out intermittently using airplanes or helicopters. Thus RF-based gamma radiation detectors can complement conventional detectors for applications that include but are not limited to emergency radiation monitoring. In the present era of advanced communication engineering domain propelled by novel antenna designs, it is indeed possible to design, fabricate and characterize RF antenna for this work on Gamma detection based on the source model explained in this work.[53,54] In short, the work reported in our study can stimulate interest in the scientific community to develop RF detectors such an antenna system with frequency of detection specific to RRL emissions arising out of interactions between Gamma radiation and water that is ubiquitously present in the vicinity of Earth's surface.

**Figure Captions**

*Figure 1. Time evolution of yield of hydrogen radical and hydrated electron for radiolysis of deaerated, pure liquid water. Reproduced with Permission from Abida Sultana et al., "Yields of primary species in the low-linear energy transfer radiolysis of water in the temperature range of 25-700°C". Phys. Chem. Chem.Phys. vol. 22, Iss. 14, (2020) P..no. 7430 -7439. Copyright @ 2020 Royal Society of Chemistry.*

*Figure 2. A schematic representation of interaction of gamma photon with water molecule revealing production of Compton electron and scattering of incident gamma photon.*

*Figure 3. A Monte Carlo Simulation Flowchart that describes the tracking of gamma photon transport including the scattering and collision events.*

*Figure 4. A Monte Carlo Simulation Flowchart that describes the electron transport including collision, excitation and ionization events.*

*Figure 5. A bird's eye view of the LabVIEW code simulating water radiolysis by Gamma Radiation.*

*Figure 6. A comparison of the population of hydrogen atoms at each L-band level with the total number of hydrogen atoms (ground state + excited) produced upon gamma radiolysis of water obtained from computational simulation.*

*Figure 7. Relative intensity of various L-band RF emissions during gamma radiolysis of water.*

*Figure 8. Variation of RF generated power with gamma photon intensity during radiolysis of water.*

*Figure 9. Spectral index plot as a function of frequency in logarithmic scale for L-band RF emission.*

*Figure 10. A plot of velocity distribution of L-band emitting hydrogen atoms.*

*Figure 11. Variation of RF generated power with absorbed dose rate for $H167\alpha$ during gamma radiolysis of water.*

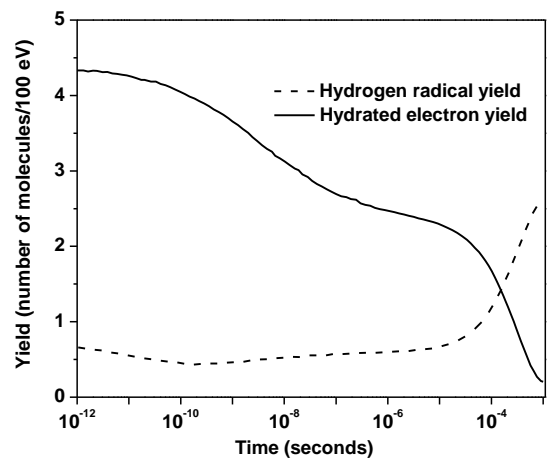

*Figure 1 of 11 Pradeep Kumar et al.*

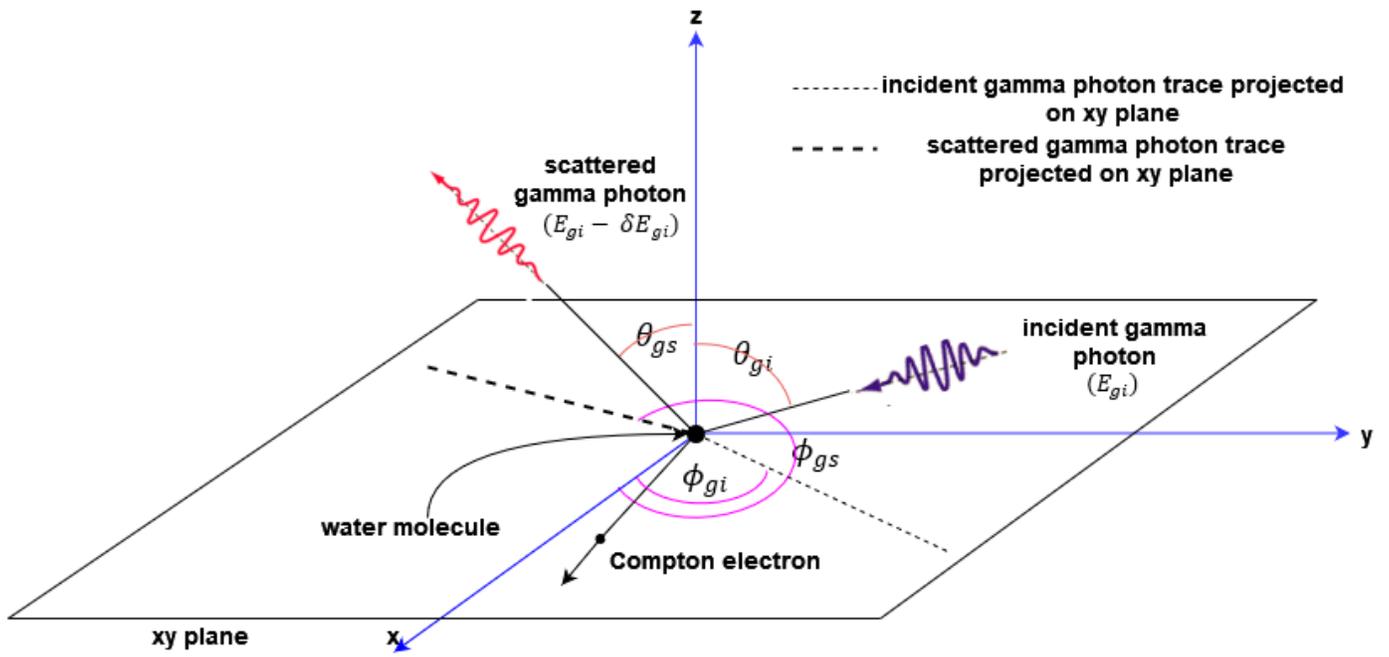

*Figure 2 of 11 Pradeep Kumar et al.*

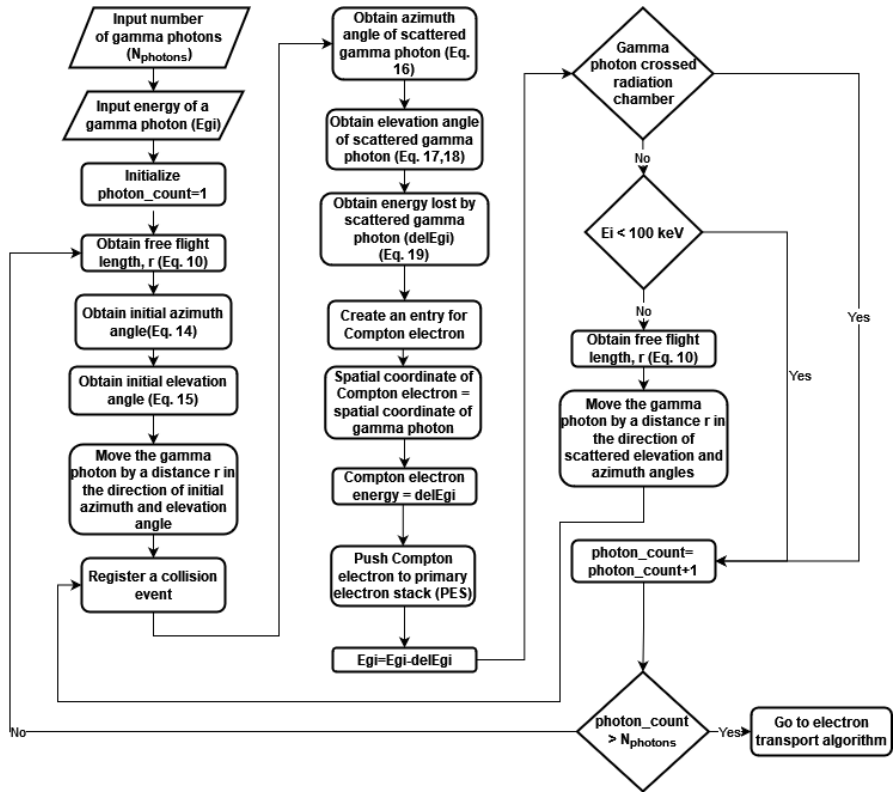

*Figure 3 of 11 Pradeep Kumar et al.*

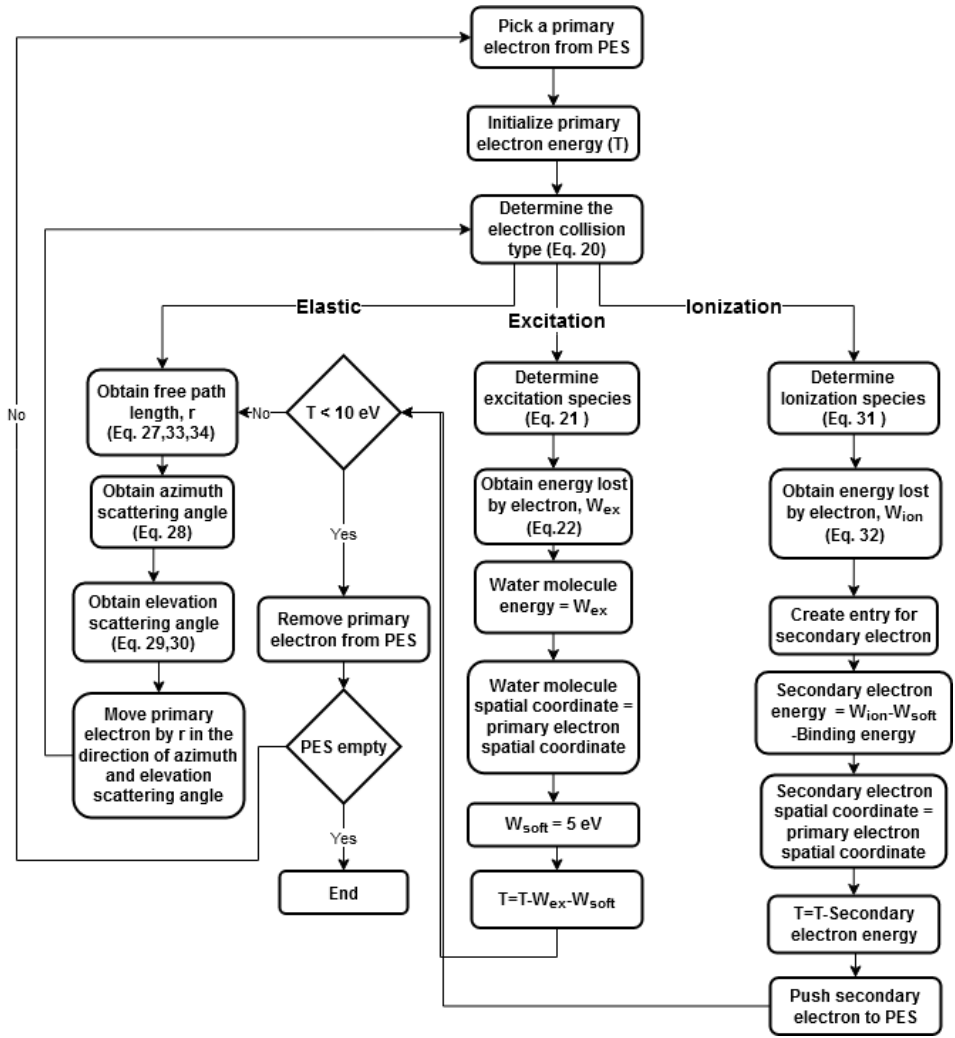

*Figure 4 of 11 Pradeep Kumar et al.*

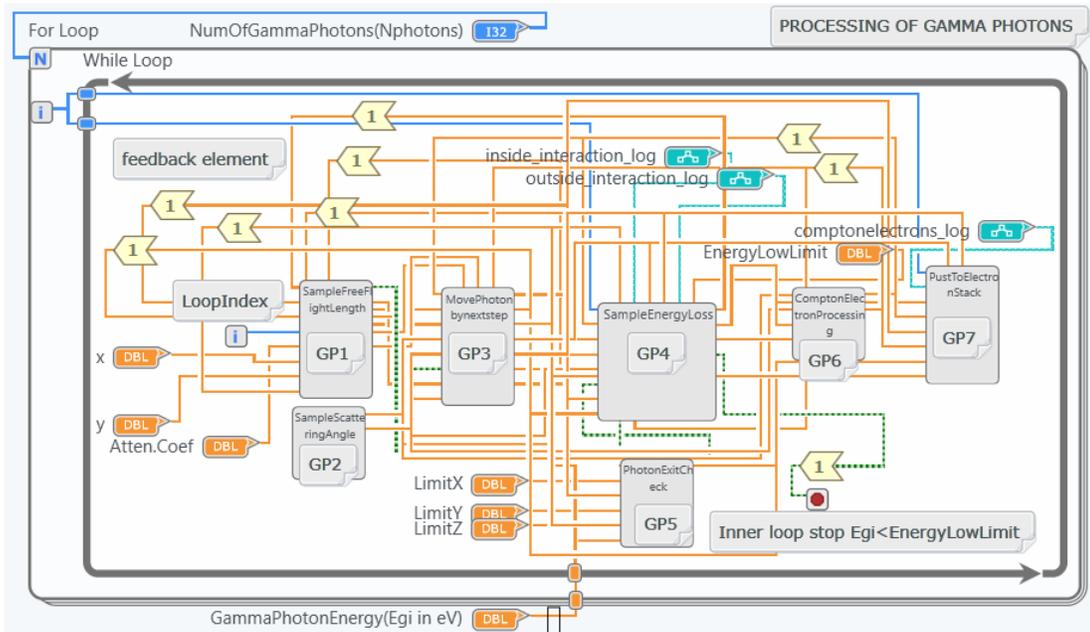
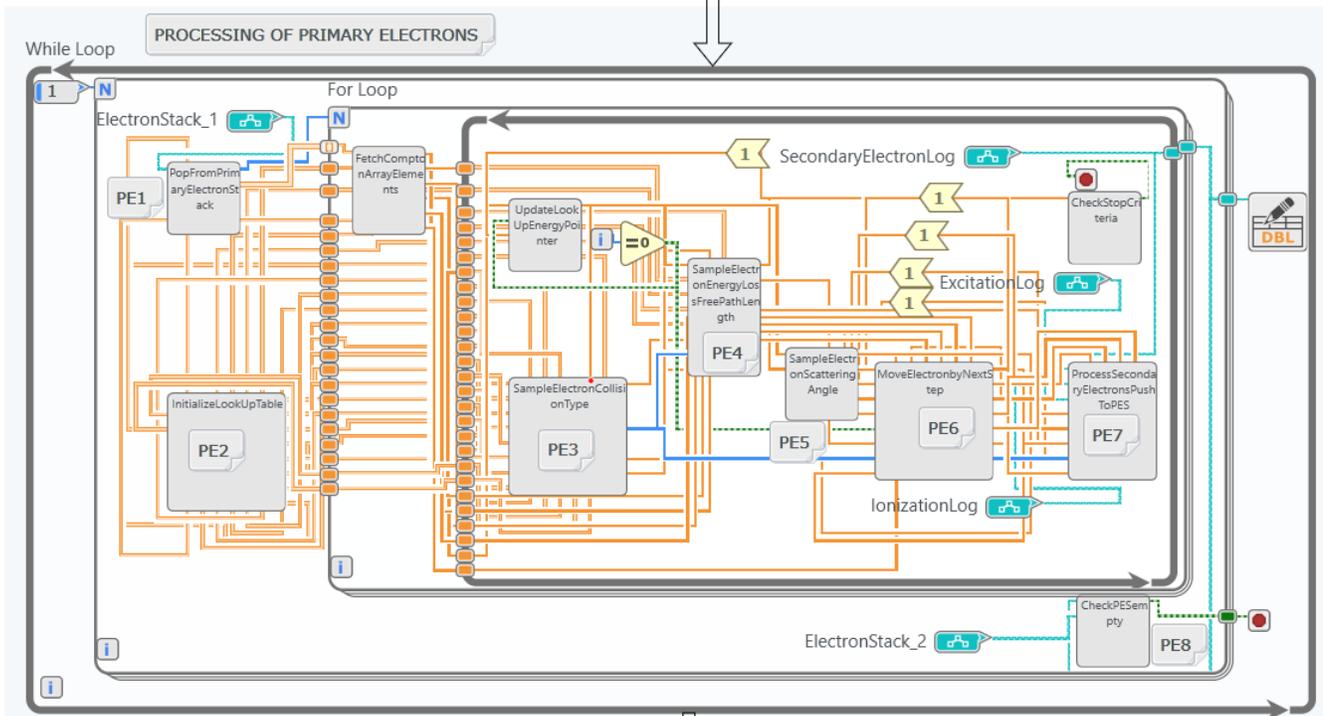
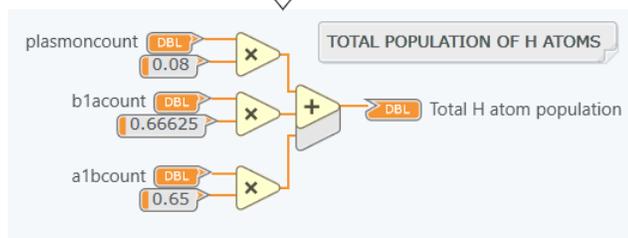

*Figure 5 of 11 Pradeep Kumar et al.*

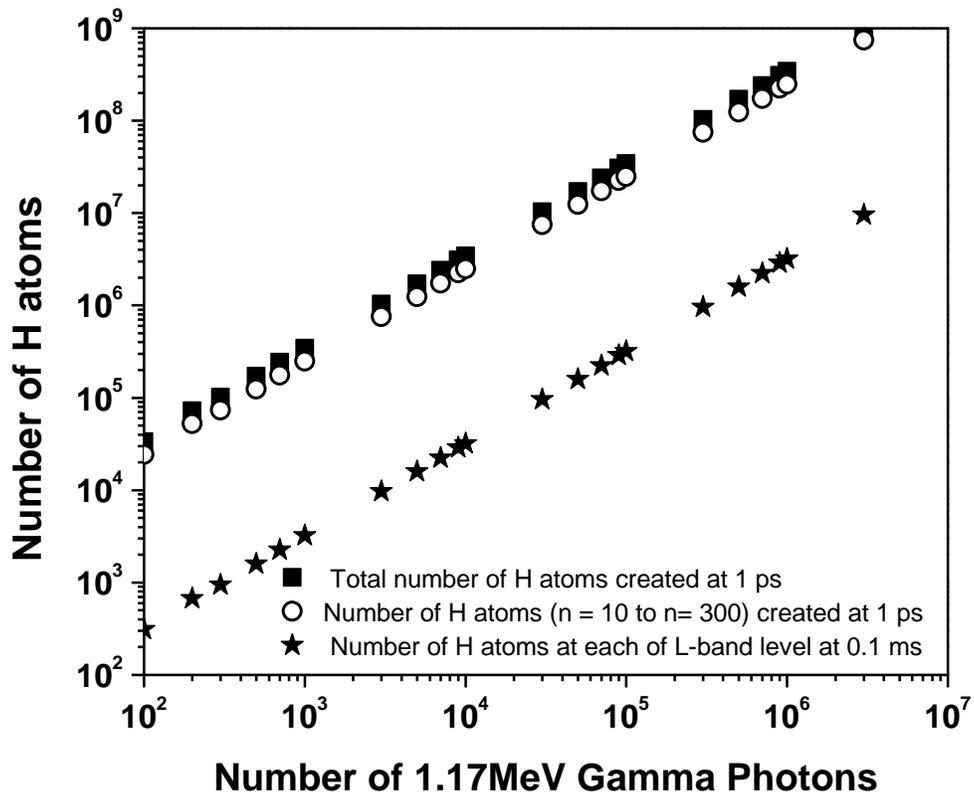

*Figure 6 of 11 Pradeep Kumar et al.*

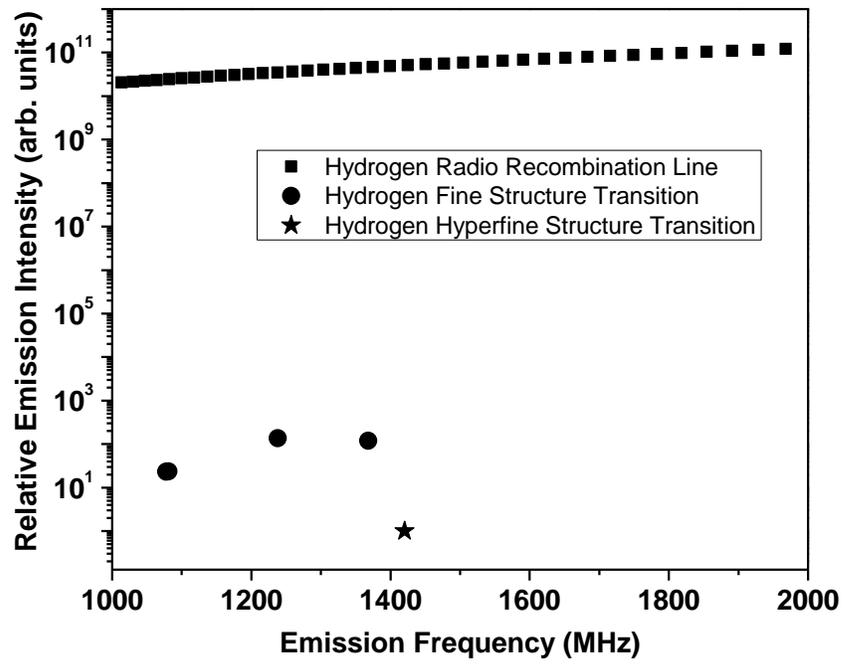

*Figure 7 of 11 Pradeep Kumar et al.*

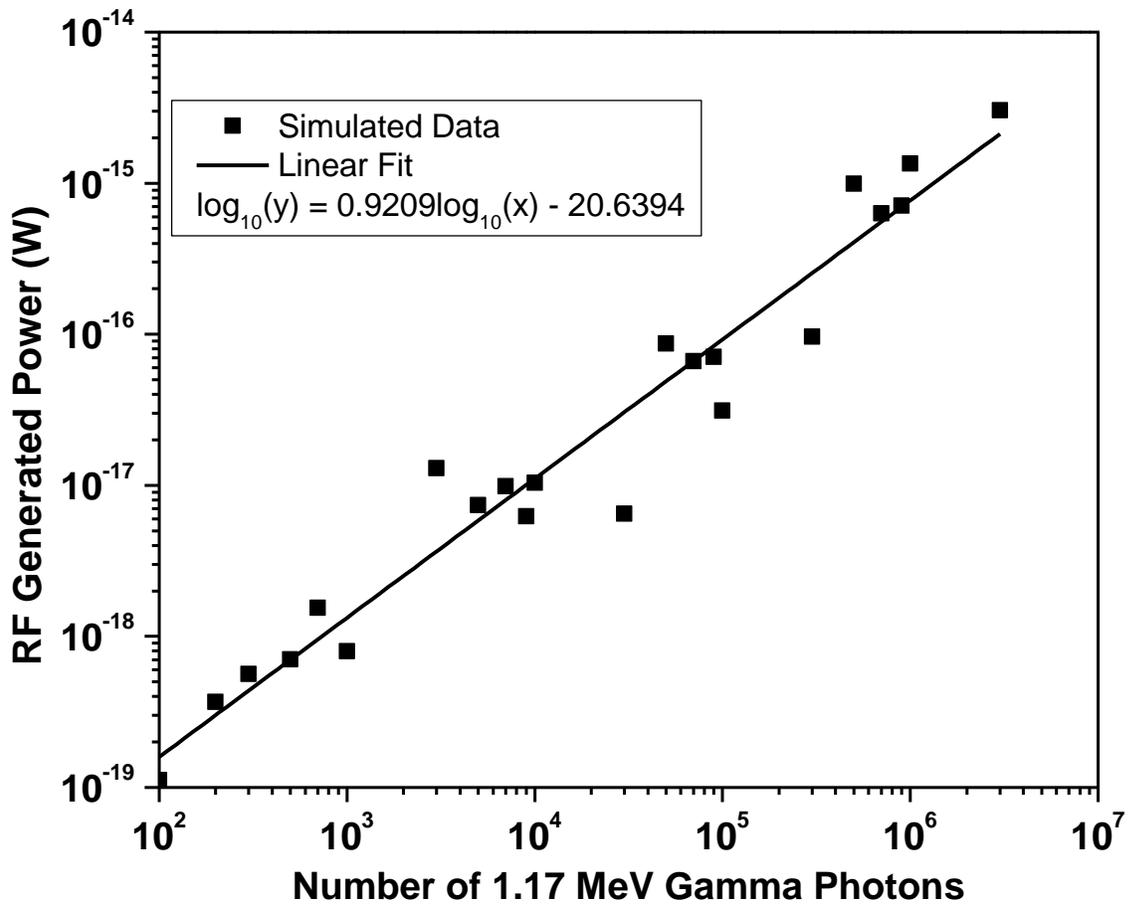

*Figure 8 of 11 Pradeep Kumar et al.*

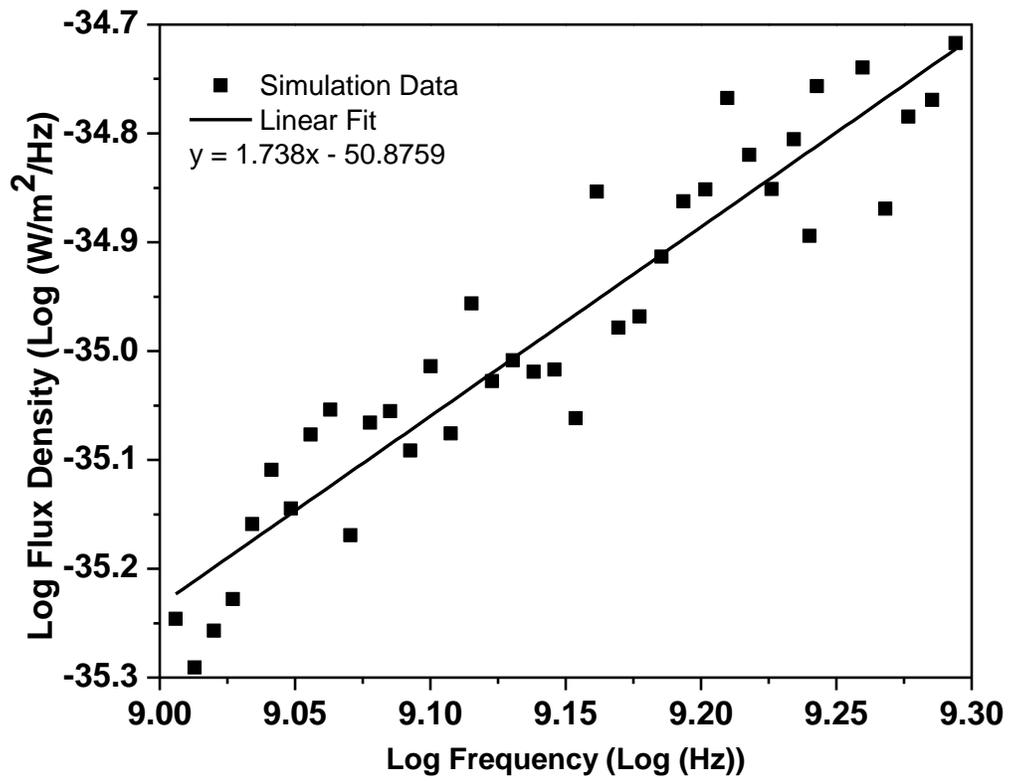

*Figure 9 of 11 Pradeep Kumar et al.,*

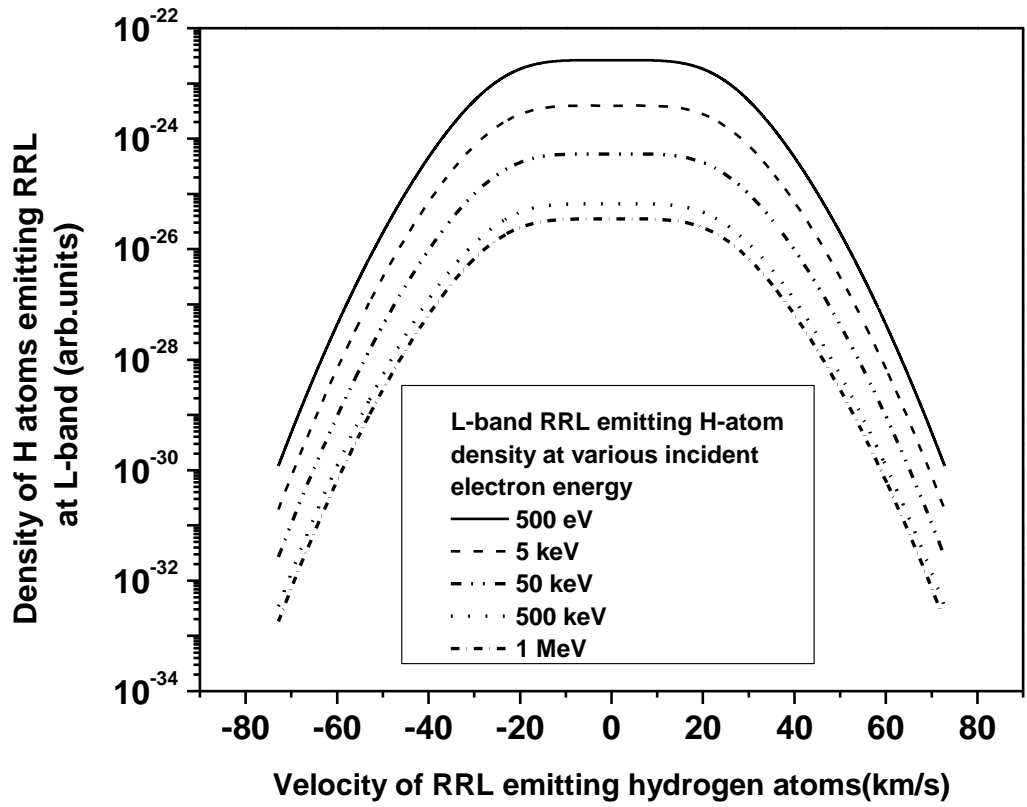

*Figure 10 of 11 Pradeep Kumar et al.,*

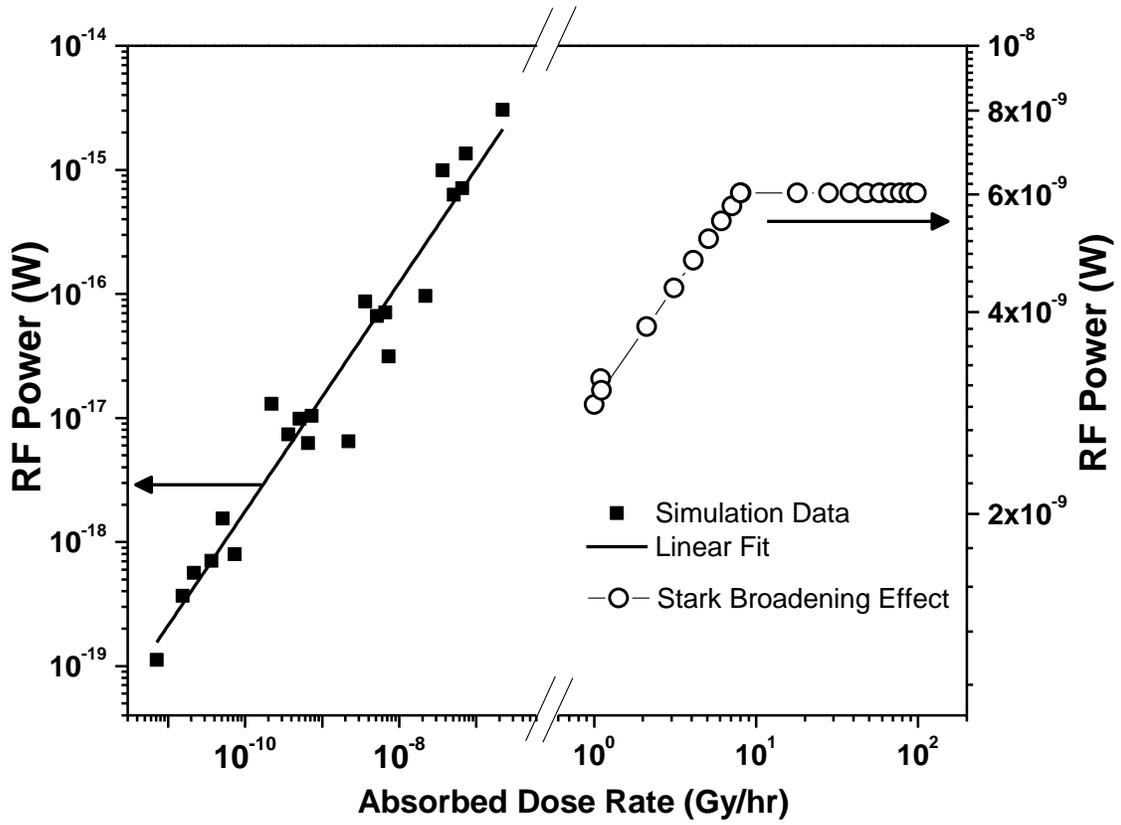

*Figure 11 of 11 Pradeep Kumar et al.,*